\documentclass[12pt,a4paper]{article}
\usepackage[utf8]{inputenc}
\usepackage[plainpages=false,pdfpagelabels=true]{hyperref}
\usepackage{amssymb,amsthm}
\usepackage[margin=1.5 in]{geometry}
\usepackage[toc,page]{appendix}
\usepackage{a4wide}  
\usepackage{amsfonts}
\usepackage{amsmath}
\usepackage{bbm}
\usepackage{ifpdf}
\newtheorem{theorem}{Theorem}[section]
\newtheorem{corollary}{Corollary}[theorem]
\newtheorem{proposition}{Proposition}
\ifpdf

\hypersetup{%
  pdftitle    = {RG-2 flow and entropy}
  pdfkeywords = {General Relativity, gravity, Ricci flow, Ads/CFT, RG flow},
  pdfauthor   = {Oscar Lasso Andino},
  plainpages  = true,
  colorlinks  = true,
  citecolor   = blue,
  urlcolor    = blue,
  linkcolor   = black
}

\newcommand{\arxiv}[1]{{\tt
\href{http://www.arXiv.org/abs/#1}{#1}}}
\newcommand{\doi}[1]{{\tt DOI:\href{http://dx.doi.org/#1}{#1}}}
\makeatletter
\@addtoreset{equation}{section}
\makeatother

\begin{document}

\begin{flushright}
\small
IFT-UAM/CSIC-18-063\\
November 09\textsuperscript{th}, 2018\\
\normalsize
\end{flushright}

\vspace{1.5cm}

\begin{center}

{\Large {\bf RG-2 flow, mass and entropy}}

\vspace{1.5cm}

\renewcommand{\thefootnote}{\alph{footnote}}
{\sl\large  \'Oscar Lasso Andino}\footnote{E-mail: {\tt oscar.lasso[at]estudiante.uam.es}}

\setcounter{footnote}{0}
\renewcommand{\thefootnote}{\arabic{footnote}}

\vspace{1.5cm}

{\it Instituto de F\'{\i}sica Te\'orica UAM/CSIC\\
C/ Nicol\'as Cabrera, 13--15,  C.U.~Cantoblanco, E-28049 Madrid, Spain}\\ \vspace{0.3cm}

\vspace{1.8cm}


{\bf Abstract}

\end{center}

\begin{quotation}
The RG-2 flow is the two-loop approximation for the world-sheet non-linear sigma model renormalization group flow. The first truncation of the flow is the well known Ricci flow, at two loops higher order curvature terms appear, changing almost completely the behaviour of the evolution equation. In this article we study the RG-2 flow  in the context of general relativity. Contrary to what happens with the Ricci flow the RG-2 flow theory has not been studied exhaustively, and from a mathematical point of view there are big differences between both flows. Considering a time symmetric foliation of an asymptotically flat spacetime, we show that the area $A$ of a closed two dimensional surface $S$ is monotonous under the RG-2 flow, refining and  extending the previous results already known for the Ricci flow. We show that the inequality that relates the area of the surface $S$ and the Hawking mass, already found for the Ricci flow, is still satisfied when we make evolve the area under the new flow. Finally, we discuss about Perelman`s $W$-entropy for the RG-2 flow and the physically motivated path towards the gradient formulation of the RG-2 flow.
\end{quotation}

\newpage
\pagestyle{plain}

\tableofcontents

\newpage


\section{Introduction}
Geometric flows have been a powerful tool for dealing with problems in the context of the so called geometric analysis, where the analytic theory of partial differential equations is used for solving geometric problems. The most famous one, the Ricci flow, has been utilized for giving a proof of the Thurston's geometrization conjecture\cite{Perelman:2006un}\cite{Perelman:2006up}\cite{Perelman:2003uq}. Originally postulated by Hamilton\cite{Hamilton:1}, the flow has been studied extensively, and has become a field of study in itself. See \cite{Rflow:1}\cite{Rflow:2}.\\
The Ricci flow is an evolution equation that deforms Riemannian metrics with the use of its Ricci tensor. This equation has many similarities with the heat equation, in the sense that it does not have memory of initial conditions and smooths out the irregularities. The appealing of this flow is the property of converging (after a finite number of "surgeries" )to canonical geometries.  
Since Perelman's work, several articles have dealt with Ricci flow applications in physics, see for example:\cite{Woolgar:2007vz}\cite{Carfora:2010iz}\cite{Headrick:2006ti}. This, occurred in part, because of Perelman's acknowledging  about his inspiration on the theory of the non-linear $\sigma$-model. That work gave another clue for relating the Ricci flow with the renormalization group flow for a QFT, indeed, the first approximation (one loop) for the renormalization group flow of the 2-dimensional non-linear $\sigma$-model is the Ricci flow.\\
One of the most important properties of the RG flow is its irreversibility, and it is believed that entropy is directly related to this property. The Zamolodchikov's c-theorem  \cite{Zamolodchikov:1986gt} in two dimensions affirms that  certain $c$-function (depending on the coupling constants and the renormalization scale)is monotonic under the RG flow. In particular, this $c$-function takes the same value as the central charge $c$ of a CFT corresponding to a fixed point of the RG flow.
Similarly, in three dimensions, the F-theorem\cite{Casini:2012ei} states that a $\mathcal{F}$-function is monotonic under the RG flow. This function takes the same value as the sphere free energy of the CFT corresponding to each fixed point of the RG flow. An analogous result can be found in four dimensions, it is called the a-therem\cite{Komargodski:2011xv}\cite{Casini:2017vbe}.\\
All these results imply that the  number of effective degrees of freedom monotonically decreases along RG flows -in the theory space of QFT-.  Usually, the monotonicity of the $c$-functions is a consequence of the strong subaditivity  and Lorentz invariance of the QFT.
This  kind of assertions suggest that  studying the monotonicity, under some flow, of an appropriate quantity could give us a lot of information. \\
The previously cited monotonicity results are stated for a CFT. However, if the RG flow corresponds to the quantization of a purely geometrical theory, such as the non-linear $\sigma$-model, there should be similar results, although the techniques for dealing with this kind of problems will be different. We can get some inspiration from other geometrical monotonicity statements. Particularly interesting are the monotonicity results under the Inverse Mean Curvature Flow (IMCF). Using this flow, geometric entropy bounds have been determined, being the proof of the Riemannian Penrose inequality a prominent example. In this proof, one of the pivotal findings was the monotonicity of the Hawking mass under the IMCF, and with this at hand, mathematicians were able to show the existence of the flow and constructed its weak formulation, see \cite{Jang:Wald}\cite{Bray:2003ns}\cite{Huisken:Ilmanen}\cite{Bray}\cite{Frauendiener:2001uf}. For a review and applications of the Penrose-type inequalities see \cite{Bray:2013owa}.\\
The Penrose inequality has not been proved in full generality, a proof of the inequality could give (maybe indirect) support to the weak cosmic censorship conjecture. Moreover, a generalization to AdS spaces has not been done, although a recent result \cite{Fischetti:2016fbh} shows, using the IMCF, that the renormalized entanglement entropy of any region of a CFT is bounded from above by a weighted local energy density. They found a ‘subregion’ Penrose inequality for asymptotically locally AdS spacetimes.\\

If we consider the two loop approximation for the renormalization group flow of the 2-dimensional non-linear $\sigma$-model we obtain the RG-2 flow. This RG-2 is quite more complicated than the Ricci flow and it has not been broadly studied. \\
In this article we want to study the evolution, under the RG-2 flow, of geometrical quantities  in the context of general relativity. Using the RG-2 flow we will be able to prove geometric inequalities in three spatial dimensions.  In order to do so, we built a 3-dimensional Riemannian manifold from a 4-dimensional spacetime manifold, by taking a time symmetric  foliation. 

In the proof of the Riemannian Penrose inequality, it is assumed the existence of the IMCF, an extrinsic flow that makes evolve a surface considering it as embedded in a bigger space. They take a time symmetric foliation of spacetime and prove that the Hawking mass is monotonic under the flow. We can address the same problem, but instead of using  an extrinsic flow, we could chose an intrinsic flow such as the RG-2 flow. The principal reasons behind the choice are two:  there is the hope that the singularities appearing in the IMCF disappear because of the diffusive character of the flow. On the other side, if there is going to be a proof of the full Penrose inequality, it has to rely in the use of an intrinsic flow over a Lorentzian manifold. Therefore, studying the evolution of geometrical quantities, in the context of GR, using intrinsic flows is the first step towards the study of that inequality.

In this context, the Ricci flow is a good candidate. However, we would like to have a flow that captures the higher curvature effects near the singularities, here the RG-2 flow is more suitable for our purposes. We will consider the RG-2 flow non perturbatively.\footnote{In the study of non-linear $\sigma$-models it is known that near points of higher curvature the Ricci flow is no longer a good approximation for the RG flow, and therefore, it should be replaced by a geometric flow involving more fields or higher curvature terms. The former is an active field of mathematics, and it could help to address problems such as the charged Penrose inequality \cite{Khuri:2014wqa}, and the latter is the intuitive way of generalizing to higher curvatures, which depending on the curvature, it could behave very different compared with the case for the Ricci flow. However,  we do not know how good the approximation is when curvatures are high. There is not a mathematically rigorous quantization of the non-linear sigma model, therefore, we cannot give a measure of the error when using Ricci flow as an approximation to the full flow. If we think about the RG-2 flow in a perturbative scenario, it will give an error estimation between the full flow and the Ricci flow approximation, at least qualitatively.
Moreover, in some field theories  it has been shown that it is enough to consider the RG-2 flow approximation of the RG flow to confirm the existence of a continuum limit.\cite{Lott:1986}}


We want to study the evolution of the area of a compact surface, that  in some cases, it can be identified with the area of an outermost horizon, and therefore, the entropy of a black hole. On the way we will see if the new flow gives a hint for a new proof of the Riemannian Penrose inequality. Moreover, we want to see if a certain type of Perelman entropy is monotonous under the RG-2 flow. With the results at hand we would like to have a clue for the gradient formulation of the RG-2 flow, which is an open mathematical problem. Since the RG-2 flow is a generalization of the Ricci flow, which includes higher order corrections in the curvature, part of our work will represent a generalization of the results already obtained for the Ricci flow.\\
The generalization is not direct, we rely on many mathematical theorems, but contrary to what happens with the Ricci flow there are not so many results about the controlled behaviour of the RG-2 flow. This flow is not weakly parabolic, and therefore, the existence of solutions is not guaranteed for all times. Moreover, there are results for the existence of the so called eternal solutions only in two dimensions, see \cite{Oliynyk:2009rh}.

In section 2 we give a review of the RG-2 flow, we focus in the evolution of a fixed constant curvature metric to see how the existence of short time solutions is guaranteed. In section 3 we study the evolution of the Hawking mass and the evolution of the area of a closed surface under RG-2 flow. We will always compare with the spherically symmetric case. Here we derive inequalities that relate the rate of change of the area of a compact surface, the Hawking mass and the curvature scalar.  In section 4 we conclude with the discussion of the results and general remarks. In the appendix we give a very short introduction to the gradient formulation of the Ricci flow and the geometrical interpretation of the Perelman's $\mathcal{W}$-entropy, giving some hints about the Gradient formulation for the RG-2 flow.
\section{The RG-2 flow}
\label{sec-generic}
Here we provide a brief introduction to the RG-2 flow and the non-linear sigma model. We will focus on compact Riemannian manifolds, and then, the results will be extended to asymptotically flat manifolds. We want to use the results listed here in a GR scenario. However, a spacetime is a Lorentzian manifold, and therefore we have to take a foliation of spacetime, then the metric induced on a spatial slice of the foliation is going to be Riemannian\footnote{We will take a time symmetric foliation of spacetime and the metric that evolves under the flow is going to be the induced metric in a 3-space. See section 3.1}. We have to remember that the RG-2 flow theorems, that let us control the behaviour of the any evolution,  are stated for 3-dimensional Riemannian manifolds not necessarily coming from a foliation of a spacetime. Thus, we have to adapt the results to the case where the 3-dimensional metric is an induced metric from a spacetime metric. Before doing so, let us see the general case.

Let $(M^{n},g)$ be a Riemannian manifold  and let $\Omega$  be a two dimensional Riemannian surface with metric $\Gamma$. We define a smooth map 
$\phi:\Omega \rightarrow M^{n}$ between the surface $(\Omega,\Gamma)$ and the Riemannian manifold $(M^{n},g)$. We also define a functional $S$ :

\begin{equation}\label{sigmamodel:1}
S(\phi,g)= \frac{1}{\alpha}\int_{\Sigma}\sqrt{-|\Gamma|} g_{ij}(\phi(x))\partial^{\mu}\phi^{i}(x)\partial^{\nu}\phi^{j}(x)\Gamma_{\mu\nu}d\sigma.
\end{equation}

The functional (\ref{sigmamodel:1}) is the general bosonic  nonlinear $\sigma$-model action. Here, $0<\alpha=\frac{T}{2}$, where $T$ is the tension of a bosonic string. $(\Omega,\Gamma)$  is usually called  the world-sheet and $(M^{n},g)$ the target space.\\ 
This is the most simple non-linear sigma model. It is usually assumed that both, the target Riemannian manifold $(M^{n},g)$ and the world-sheet surface $(\Omega,\Gamma)$ are compact, orientable and without boundary. The scalar fields  $\phi^{i}$  are taken as elements of the space of smooth maps $C^{\infty}(\Omega,M^{n})$.\\
In the process of perturbatively quantizing the action (\ref{sigmamodel:1}) it is necessary to introduce a momentum cuttof $\Lambda$. With this regularization the divergent section of the theory is confined to a term with the logaritmic energy $\log\Lambda$ which multiplies the Ricci tensor $R_{ij}$, then the target space metric becomes dependent of the cutoff. This cutoff $\Lambda$ parametrizes a family of quantum field theories as the distance scale changes from $\frac{1}{\Lambda}$ to $\frac{1}{\Lambda'}$. 
Setting $\lambda:=-\ln (\frac{\Lambda}{\Lambda'})$ the renormalization group flow for the theory becomes:

\begin{equation}\label{renequa:2}
\frac{\partial g_{ij}}{\partial \lambda}=-\beta_{ij}(g).
\end{equation}

The beta functions $\beta_{ij}$ are given perturbatively, and  if we want to calculate higher order terms for the beta functions we have to consider Feynamnn diagrams at more loops. See \cite{Ketov:2000} for a good introduction to the quantum non-linear $\sigma$-model. For example, at three loops, the $\beta_{ij}$-function for the non-linear bosonic sigma model can be written \cite{Jack:1989vp}:
\begin{equation}
\beta_{ij}=\alpha\beta_{ij}^{(1)}+\alpha^{2}\beta_{ij}^{(2)}+\alpha^{3}\beta_{ij}^{(3)},
\end{equation}
where
\begin{eqnarray}
\beta_{ij}^{(1)}&=&R_{ij},\nonumber\\
\beta_{ij}^{(2)}&=&\frac{1}{2}R_{iklm}R_{j}^{\,\,klm},\\
\beta_{ij}^{(3)}&=&\frac{1}{8}\nabla_{p}R_{iklm}\nabla_{p}R_{j}^{\,\,klm}-\frac{1}{16}\nabla_{i}R_{klmp}\nabla_{j}R^{klmp}+\frac{1}{2}R_{klmp}R_{i}^{\,\,mlr}R_{j\,\,\,\,r}^{\,\,kp}-\frac{3}{8}R_{iklj}R^{kspr}R^{l}_{\,\,spr}.\nonumber
\end{eqnarray}

In this work we will consider the 2-loop beta functions, thus the right hand side of the flow equation (\ref{renequa:2}) becomes:

\begin{equation}
\beta_{ij}(g)=\alpha R_{ij}+\frac{\alpha^{2}}{2}R_{iklm}R_{j}^{\,\,klm}+O(\alpha^{3}),
\end{equation}

then, the renormalization group flow equation (\ref{renequa:2}) is, after a reescalation in $\lambda$:

\begin{equation}\label{renequanew:3}
\frac{\partial g_{ij}}{\partial \lambda}=-2R_{ij}-\frac{\alpha}{2}R_{iklm}R_{j}^{\,\,klm}.
\end{equation}
The flow defined by  (\ref{renequanew:3}) is called the \textbf{RG-2 flow} and represents the first next to leading order contribution to the RG flow of the non-linear sigma model, the RG-1 (1-loop flow) is the Ricci flow \cite{Friedan:1980jf}.  \\

In this article, since we will work in the context of GR, we will use the RG-2 flow non perturbatively, meaning that in general we do not put any restriction over $\alpha$ apart from being non-negative, and every ``correction" in $\alpha$ will change the old flow into another flow. We will talk about higher loop corrections in the sense that we have a flow with more $\alpha$ terms.\\ 
It is known \cite{Gimre:etal} that when $Rm_{ij}^2$ reaches the value near to $\frac{2}{\alpha}$ the solutions of the RG-2 flow behave as the ones for the Ricci flow, but the families of solutions with large negative curvature behave very differently.\\

There are some special zones where the flow is weakly parabolic. For example, it has been proven  that the RG-2 flow is weakly parabolic in any dimension, only in the zones where $1+\alpha K_{ij}>0$ ,. \cite{Gimre:2014jka}, where the $K_{ij}$ are the sectional curvatures of the Riemannian manifold. This is the result that we will be using through the entire article.\\
The evolution under RG-2 flow of the volume element $d\mu$  is given by  \cite{Gimre:etal2}
\begin{equation}\label{vol}
\frac{\partial d\mu}{\partial\lambda}=\left(-R-\frac{\alpha}{4}|R_{m}|^{2}\right)d\mu,
\end{equation}
here $R$ denotes the scalar curvature of the manifold. According to (\ref{vol}), if $R>0$ then the volume element decreases, but when $R<0$ the  rate of change can be positive or negative depending on the curvature and the factor $\frac{2}{\alpha}$. Since the initial formulation of General Relativity requires it we will be interested in the case $R>0$.

\subsection{Initial data}\label{idata}

According to the initial value formulation of general relativity, we can choose a spacelike hypersurface $\Sigma$, which represents a foliation (an instant) of spacetime, whose 3-dimensional metric (induced form the spacetime metric) will describe the intrinsic geometry of $\Sigma$. Then, our flow is an intrinsic flow in that 3-dimensional space. More specifically, the spacetime metric  $g_{\alpha\beta}$ induces a metric $h_{ab}$ over $\Sigma$, and as usual , it is given by
\begin{equation}
h_{ab}=g_{\alpha\beta}e^{\alpha}_{a}e^{\beta}_{b},
\end{equation}
where $e^{\alpha}_{a}$ and $e^{\beta}_{b}$ are tangent vectors to the surface.\\
In \cite{Gimre:2014jka} the authors show that there is a zone where the operator corresponding to the RG-2 flow is weakly elliptic, and therefore the  DeTurck trick can be applied, proving that the Cauchy problem of the evolution of a three dimensional, smooth, compact, Riemannian manifold is equivalent to a Cauchy problem for a strictly parabolic operator, only in the zones where $1+\alpha K_{ab}>0$.\\

We will consider an asymptotically flat Riemannian manifold $(M,h_{ab}(\lambda))$ with dimension three (a,b =1,2,3). The RG-2 flow given in eq.(\ref{renequanew:3}) will make evolve the three dimensional metric $h_{ab}$ giving as a result a parametric family of metrics. 
The RG-2 flow eq.(\ref{renequanew:3}) with a DeTurck term becomes:
\begin{equation}\label{RG:2turck}
\frac{\partial h_{ab}}{\partial\lambda}=-2R_{ab}-\frac{\alpha}{2}R_{apqs}R_{b}^{\,\,\,pqs}+\nabla_{a}V_{b}+\nabla_{b}V_{a},
\end{equation}
where $V_{a}$ is the DeTurck vector, which will be required to decay at infinity.
Moreover, we will call the RGo-2 flow to the following \begin{equation}
\frac{\partial h_{ab}}{\partial\lambda}=-\frac{\alpha}{2}R_{apqs}R_{b}^{\,\,\,pqs}+\nabla_{a}V_{b}+\nabla_{b}V_{a},
\end{equation}
which is parabolic in the zones where $\alpha K_{ab}>0$.\\
We have to be sure that our flow evolves the geometric quantities of interest in a controlled way. This formulation incorporates the two symmetric tensors $h_{ab}$ and $K_{ab}$ defined in the space-like surface $\Sigma$. Here $K_{ab}$ is the extrinsic curvature of $\Sigma$. They are constrained by the Einstein equations and by the two following relations:
\begin{align}
\nabla_{b}(K^{ab}-h^{ab}K)&= 8\pi J^{b}\label{GaussC1}\\
_{3}R+K^{2}-K_{ab}K^{ab}&=16\pi\rho \label{GaussC2},
\end{align}
where $J^{b}$ is the matter current and $\rho$ is the matter density. There are evolution equations for $h_{ab}$ and $K_{ab}$, but since we are interested in the RG-2 evolution in a fixed foliation of the spacetime we will not consider such equations.
We consider a time symmetric initial data $(\Sigma,h_{ab})$, which means that the extrinsic curvature has to be set to zero, thus $(\Sigma,h_{ab})$ is totally geodesic. Setting to zero the matter current we arrive at the only condition that has to be satisfied:
\begin{equation}\label{Rpositive}
R=16\pi\rho.
\end{equation}
Usually, energy conditions imply the non-negativeness of the matter density, which in its turn -because of equation (\ref{Rpositive})- implies the non-negativeness of $R$. It is desirable that our flow preserves this property of the scalar curvature $R$. There are some results that help to control the evolution of the curvature under the RG-2 flow, although the most general case has not been proved there are some results that show that our flow will maintain the sign of the scalar curvature \cite{Gimre:2014jka}\cite{Oliynyk:2009rh}\\

\subsection{Constant curvature solutions, fixed points and solitons}\label{solutions}
In the previous section we cited some results that ensure the existence of solutions, but if we want RG-2 flow to be useful in Physics we should be able to characterize at least some simple solutions. To get a taste about the evolution under the RG-2 flow we will study a simple example. The simplest solutions (after the fixed points) are the constant curvature solutions. There is a mathematical proof for the existence of eternal solutions in two dimensions \cite{Oliynyk:2006nr},and although the most general case has not been studied there are some results that can help us for understanding the behaviour of our flow. The constant curvature case is studied in \cite{Gimre:etal}, see also \cite{Gimre:etal2}. There is a simple characterization for metrics satisfying  $g(\lambda)=f(\lambda)g_{K}$, where $g_{K}$ is a fixed constant curvature $K$ metric. In $d-$dimensions the function $f(\lambda)$ has to be of the form:
\begin{equation}\label{solflow:1}
f(\lambda)=1-2K(d-1)\lambda+\frac{\alpha K}{2}\ln\left|\frac{2\phi(\lambda)+\alpha K}{2+\alpha K}\right|,
\end{equation}
and it will develop a singularity in a finite time $T$ given by the expression
\begin{equation}\label{ctime:1}
T=\frac{1}{2K(d-1)}+\frac{\alpha}{4(d-1)}\ln \left|\frac{\alpha K}{2+\alpha K}\right|.
\end{equation}

The collapsing time $T$ given in (\ref{ctime:1}) determines the finite time when the metric $g(\lambda)$ will develop a singularity, and it  works only for the case when $K$ is not equal to zero or $-\frac{2}{\alpha}$. 
\bigskip

As an example, let us consider a metric given by
\begin{equation}\label{Robertson}
ds_{3}^{2}=a^{2}(\lambda)\left(\frac{dr^{2}}{1-K r^{2}}+r^{2}d\theta^{2}+r^{2}sin^{2}(\theta)d\phi^{2} \right),
\end{equation}
where $a$ is constant in time. In this case the condition $g(\lambda)=a(\lambda)g_{K}$ is satisfied for 
$g_{K}$ given by
\begin{equation}
g_{K}=\frac{dr^{2}}{1-K r^{2}}+r^{2}d\theta^{2}+r^{2}sin^{2}(\theta)d\phi^{2}.
\end{equation}
Using eq.(\ref{solflow:1})  we found that $a(\lambda)$ is given by  
\begin{equation}\label{aeq}
a(\lambda)=-\frac{K}{2}\left(1+W_{-1}\left(-\frac{(2+K)}{K}e^{-1-\frac{2}{K}+8\lambda} \right)\right),
\end{equation} 
where $W_{-1}$ is one of the branches of the Lambert function\footnote{The Lambert W is the inverse function of $f(W)=We^{W}$ and has two branches $W_{0}$ and $W_{-1}$}. \\

The conditions (\ref{solflow:1}) and (\ref{ctime:1}) work for any Riemannian metric. However, if we want to use the in a GR scenario we have to take into account the restrictions imposed by the projected Enstein Equations. In this particular case, if we assume that (\ref{Robertson}) is a spacetime induced metric over a spacelike hypersurface then the restriction (\ref{GaussC1}) is trivially satisfied\footnote{Because of isotropy and homogeneity we have that $\rho=cte$ and $j^{b}=0$. Moreover $K_{ab}=\frac{1}{2}\left(\frac{\dot{a}}{a}\right)h_{ab}$, where the overdot denotes the time derivative.  The restriction (\ref{GaussC2}) will give a Friedmann type equation, which in its turn leads to the constancy of $\left(\frac{\dot{a}}{a}\right)$. }.The 3-dimensional metric (\ref{Robertson}) will correspond to the time symmetric foliation of a Friedman-Robertson-Walker (FRW) type spacetime \footnote{If $\dot{a}=0$ we have that $K_{ab}=0$ which implies that we have a time symmetric foliation.}.\\
Depending on the values of $K$ and $\alpha$ the flow will have different particularities.
When $K>0$ the sphere of constant curvature shrinks to a point in a time given by (\ref{ctime:1}). Interestingly, since $\frac{\alpha K}{2+\alpha K}>0$ for $\alpha>0$ the time to singularity is smaller than the corresponding time for the same geometry under Ricci flow. The time for collapsing depends directly on the curvature, if curvature is very big, then the time for collapsing  is very small. In the limit when $K$ goes to infinity the time for collapsing goes to zero.\\
In physical applications $\alpha$ is always positive (it corresponds to the inverse of the length of a string). When $K<0$ and $2+\alpha K<0$ the curvature $K$ is very big and negative, then the metric collapses to a point in finite time. Here we have an example of a space which is not singular under Ricci flow -under this flow it expands forever-, but under RG-2 it develops singularities. We will see that this fact changes completely the behaviour of geometric quantities under the RG-2 flow.
For the case when $K<0$ and $K>-\frac{\alpha}{2}$, $K$ can be very small and therefore the geometry behaves very similar as under Ricci flow. Finally,  when $2+\alpha K=0$ and $K<0$ we have a fixed point.\\
As we have seen there are some regimes where the RG-2 flow behaves as the Ricci flow, in any case, when we set $\alpha$ to zero we should recover the results of Ricci flow.
As we pointed out the simplest solutions are the fixed points of the flow, since we want to relate them with the gradient formulation of the flow we will give here a brief description. \\
A fixed point of the equation (\ref{renequanew:3}) is a solution that does not change in time. Some of the fixed points of the RG-2 flow are well know: metrics with negative Ricci curvature. In 3-dimensions, the fixed point equation can be written:
\begin{equation}\label{fixedpointdec}
R_{ab}=-\frac{\alpha}{2}\left(-R^{s}_{a}R_{sb}+RR_{ab}+|Rc|^{2}h_{ab}-\frac{R^2}{2}h_{ab}\right).
\end{equation}  
Diagonalizing the Ricci tensor as a function of an orthonormal basis, it has been shown \cite{Gimre:etal2} that there are four solutions to the equation (\ref{fixedpointdec}). Three of those solutions  have been identified, they are: ${\rm I\!R}^3$,${\rm I\!H}^3$ and ${\rm I\!H}^{2}\times{\rm I\!R}$. If we want to use geometries more complex than those of the fixed points we should be able to study solutions that evolve self-similarly. Those kind of solutions are called solitons, they are very important and have not been studied in depth \cite{Wears:2016}. For a steady gradient soliton $h_{0}$ that at $\lambda=0$ satisfies  
\begin{equation}\label{soliton:1}
_{0}R_{ab}+\frac{\alpha}{2}(_{0}Rm_{ab}^{2})+_{0}\nabla_{a0}\nabla_{b} f_{0}=0,
\end{equation}
for a given function $f_{0}=f(0)$, there exist a gradient soliton $h_{\lambda}$ (with gradient function $f_{\lambda}$) whose initial data is given by $h_{0}$ and $f_{0}$.
If the manifold $M$ where both solitons are defined has constant curvature then, there is always a solution to the RG-2 flow equation. Although there is not known a gradient formulation for the RG-2 flow we will argue in favour of this, and also we will give some clues about the deformation of the flow  in order to get some hints about this gradient  formulation, see appendix A.

\section{RG-2 flow and entropy}
\label{energy-entropy}
We have introduced the RG-2 flow with all results stated for compact manifolds, but if we are interested in the mass and entropy of black holes for example,  we should consider non compact manifolds. Moreover, want to know how geometric quantities evolve when the induced metric $h_{ab}$ evolves with the RG-2 flow. In this section we particularize to the case where the evolving Riemannian metric is a spacetime metric induced over a space-like hypersurface. See section \ref{idata}. Here we present our main results.

\subsection{Local geometric quantities}

Let $S$ be a closed surface on $\Sigma$, we call $\gamma_{ij}$ to the induced metric on $S$, $\mathcal{R}$ is the scalar curvature of $S$ and $K$ is the trace of the extrinsic curvature $K_{ab}$. We are interested in local quantities (mass) and its evolution.

For defining a local mass we should be able to find a functional $\mathcal{L}$ depending on the metric and its derivatives such that
\begin{equation}
M=\oint_{S}\mathcal{L}(g_{\mu\nu}\partial{g_{\mu\nu}})dS.
\end{equation}
Nobody knows a $\mathcal{L}$ that makes $M$ to satisfy  all properties required for being a local mass.
There are some good definitions of quasi-local mass, meaning that they satisfy almost all properties required.  One remarkable example is the Geroch mass which has been used by Geroch for giving a proof of the positivity of ADM mass. This Geroch mass is a non-Lorentz invariant modification of the Hawking mass. We will study the evolution of the Hawking mass, which is related to the energy, and is defined for any closed surface $S$. It has the interpretation of the mass contained within the surface $S$. The Hawking mass is monotonic along the inverse mean curvature flow \cite{Jang:Wald}, this fact has been used to give a proof of the Riemannian Penrose inequality for a single black hole by  Huisken and Ilmanen \cite{Huisken:Ilmanen}, later on Bray did it for the multiple black hole case.\cite{Bray}.
Moreover, the Hawking mass is monotonic on asymptotically null or hyperbolic hypersurfaces\cite{Sauter} or in a general spacetime \cite{Bray:2013owa}\cite{Frauendiener:2001uf}. All those properties make the Hawking mass a very convenient quantity for making evolve under RG-2 flow. \\
Given a closed surface $S$, its Hawking mass is given by
\begin{equation}\label{Hawking:1}
M_{H}=\sqrt{\frac{A}{16\pi}}\left(1+\frac{1}{16}\oint_{S}\theta_{+}\theta_{-}dS\right),
\end{equation}
where the $\theta_{\pm}$ are called the null expansions of $S$ and $A$ is the area of $S$, which is given by
\begin{equation}\label{area:1}
A(S)=\int_{S}dA=\int_{S}d^{2}x\sqrt{\gamma}.
\end{equation}
The equation (\ref{Hawking:1}) can be written as
\begin{equation}
M_{H}(S)=\frac{1}{16}\left(\frac{A(S)}{16 \pi^{3}}\right)^{\frac{1}{2}}\int_{S}dA(2\mathcal{R}-K^{2}).
\end{equation}

It is customary to use the compactness $C(S)$, a dimensionless quantity given by
\begin{equation}\label{compactness}
\mathcal{C(S)}=\int_{S}dA(2\mathcal{R}-K^{2}).
\end{equation}
The compactness have been used before in the proof of the Riemannian Penrose inequality and in the proof of the positive mass theorem, it goes to zero when $S$ tends to a sphere of small radius, it also has a physical interpretation\cite{Samuel:2007ak}.

\subsection{The initial metric}
The easiest type of metrics that we can make evolve are the homogeneous spaces, in this case the non linear PDE's become ode's, and are much more easily solved. The behavior of Ricci Flow in two, three,four and five dimensional homogenous manifolds has been studied and it is mostly understood \cite{Isenberg:1992aey}\cite{Isenberg:etal}\cite{Hirschman:Bell}. Similarly, the study of homogeneous compact manifolds under RG-2 flow have been studied \cite{Gimre:etal}\cite{Glickenstein:etal2017}\cite{Isenberg:Jackson2}\cite{Glickenstein:Payne}. However, we want a little bit more freedom when applying our flow to spacetime metrics which are asymptotically flat, because in this case a homogeneous space will be flat everywhere. On the other hand, spherical symmetry helps to reduce our variables to two, namely $r$ and $\lambda$. It also allows working with non-trivial spaces. Following \cite{Poisson} and using the notation given in \cite{Samuel:2007ak} we will define two types of metrics, referring to different choices of coordinate gauge, the \textbf{a} form of the metric and the \textbf{b} form of the metric.
In the  \textbf{a} form, the initial metric of the spacelike foliation is written
\begin{equation}
ds^{2}=a(r,\lambda)dr^{2}+r^{2}d\Omega^{2}_{2},
\end{equation}
where $d\Omega^{2}_{2}=(d\theta^{2}+\sin^{2}(\theta)d\phi^{2})$. The Ricci scalar $R$ and the non-zero components of the tensors  $R_{ij}, Rm_{ij}$ are
\begin{align}
R=\frac{2}{r^{2}}-\frac{2}{r^{2}a}+\frac{2a'}{ra^{2}},\\
R_{rr}=\frac{a^{'}}{r a},\,\,\,\,\,\,Rm_{r,r}=\frac{a^{'2}}{r^{2}a^{3}},
\\
R_{\theta\theta}=1-\frac{1}{a}+\frac{ra^{'}}{2a},\,\,\,\,\,\
Rm_{\theta\theta}=\frac{2}{r^{2}}+\frac{2}{r^{2}a^{2}}-\frac{4}{r^{2}a}+\frac{a^{'2}}{2a^{4}},\\
R_{\phi\phi}=\sin^{2}(\theta)R_{\theta\theta},\,\,\,\,\,\ Rm_{\phi\phi}=\sin^{2}(\theta)R_{\theta\theta}.
\end{align}
We will use this metric for the evolution of the Hawking mass, and it can be used also, for the evolution of the ADM mass \cite{Lasso-Andino:1}. Thus, the Hawking Mass on this metric for a sphere $S$ of constant radius $r$ can be written as 
\begin{equation}
M_{H}(S)=\frac{1}{4}\sqrt{\frac{A(S)}{\pi}}\left(1-\frac{1}{a(r)}\right).
\end{equation}

When $a(r)=1$ we recover flat space and $M_{H}(S)=0$. On the other hand, if $a(r)=\left(1-\frac{2M}{r}\right)^{-1}$ and for $r>2M$ we get $M_{H}(S)=M$. Allowing a dependence of $M$ on $r$, $a(r)=\left(1-\frac{2M(r)}{r}\right)^{-1}$ we can write $M_{H}(S)=M(r)$ and then the compactness eq.(\ref{compactness}) becomes:
\begin{equation}
C(r)=32\pi\frac{M(r)}{r}.
\end{equation}
The two quantities $C(S)$ and $M_{H}(S)$ will play a crucial role in the present work.
The \textbf{a} metric is not suitable for studying apparent horizons, for this case the \textbf{b} form is used. The initial metric on this form is
\begin{equation}\label{bform}
ds^{2}=dr^{2}+b(r,\lambda)d\Omega^{2}_{2}.
\end{equation}
Similarly, the non zero components of the quantities $R_{ij}, Rm_{ij}$ and the scalar curvature $R$ are :
\begin{align}
R=-\frac{2b^{''}}{b}+\frac{b^{'2}}{2b^{2}}+\frac{2}{b},\\
R_{rr}=\frac{b^{'2}}{2b^{2}}-\frac{b^{''}}{b},\,\,\,\,\,\,\,Rm_{rr}=\frac{b^{'4}}{4b^{4}}-\frac{b^{'2}b^{''}}{b^{3}}+\frac{b^{''2}}{b^2},\\
R_{\theta\theta}=1-\frac{b^{''}}{2},\,\,\,\,\,\, Rm_{\theta\theta}=\frac{2}{b}-\frac{b^{'2}}{b^{2}}+\frac{b^{'4}}{4b^{3}}
-\frac{b^{'2}b^{''}}{2b^{2}}+\frac{b^{''2}}{2b},\\
R_{\phi\phi}=\sin^{2}(\theta)R_{\theta\theta},\,\,\,\,\,\, Rm_{\phi\phi}=\sin^{2}(\theta)R_{\theta\theta}.
\end{align}

The Hawking  mass of a sphere $S\subset\Sigma$  of constant radius $r$ on this metric is given by\cite{Samuel:2007ak}
\begin{equation}
M_{H}(r)=\frac{\sqrt{b}}{2}\left(1-\frac{b'^{2}}{4b}\right).
\end{equation}
When $b=r^{2}$ we recover flat space and then $M_{H}(r)=0$. The compactness $C(r)$ becomes
\begin{equation}
C(r)=\frac{4\pi}{b}(4 b - b^{'2}).
\end{equation}
Now we can start making geometrical quantities evolve under our flow. Depending of what we need we will be using the $\textbf{a}$-form or the $\textbf{b}$-form of the metric.

\subsection{Area of apparent horizon under RG-2 flow}\label{areaflow:1}

Let us consider a metric in the $\textbf{b}$-form, we will make evolve a fixed ($\frac{d\lambda}{dr}=0$) surface $S$ on $\Sigma$ under the RG-2 flow. Before calculating the evolution of the area we would like to know if the RG-2 flow make minimal surfaces appear. Let us assume that there is an apparent horizon in $r=r_{a}$. If $n_{i}$ is the radial, unit normal to the surface horizon then the trace of the extrinsic curvature is
\begin{equation}
K:\nabla_{a}n^{a}=\frac{b^{'}}{b}.
\end{equation}
Using the condition $K=0$ for the existence of an apparent horizon at  $r=r_{a}$,we conclude  
\begin{equation}\label{apparent}
b^{'}|_{r=r_{a}}=0.
\end{equation}
It is known that the Ricci flow cannot make appear a minimal surface if it was not present at the beginning of the evolution \cite{Samuel:2007ak}, we can explore what happens with the RG-2 flow. 
\begin{proposition}
Let us consider  an asymptotically flat spherically symmetric spacetime written in the b-form. Then trapped regions cannot spontaneously appear because of the RG-2 flow. If there is any trapped surface in the beginning, the flow will make that surface disappear.
\end{proposition}
We will adapt the proof for the Ricci flow for the case of the  RG-2 flow, although with some modification we will see that the result holds.
\begin{proof}
The  radial and angular components of the RG-2 flow given in eq.(\ref{RG:2turck}) are, 

\begin{eqnarray}
\frac{\partial h_{rr}}{\partial\lambda}=-2R_{rr}-\frac{\alpha}{2}R_{rpqs}R_{r}^{\,\,\,pqs}+2(\nabla V)_{rr},\label{RG:2r}\\
\frac{\partial h_{\theta\theta}}{\partial\lambda}=-2R_{\theta\theta}-\frac{\alpha}{2}R_{\theta pqs}R^{\,\,\,pqs}_{\theta}+2(\nabla V)_{\theta\theta},\label{RG:2t}
\end{eqnarray}

from the eq.(\ref{RG:2t}) we obtain
\begin{equation}
\frac{\partial b}{\partial\lambda}=-2+b''-\frac{\alpha}{2}\left(\frac{2}{b}-\frac{b^{'2}}{b^{2}}+\frac{b^{'4}}{4b^{3}}
-\frac{b^{'2}b^{''}}{2b^{2}}+\frac{b^{''2}}{2b}\right),
\end{equation}
taking $V_{r}=\partial_{r}f$ and using the $r$ component of the flow eq.(\ref{RG:2r}), we can take $f$ such that
\begin{equation}
f''=\frac{b^{'2}-2bb''}{b^{2}}+\left(\frac{b^{'2}}{2b^{2}}-\frac{b^{''}}{b}\right)^{2}.
\end{equation}
The eq.(\ref{RG:2t}) now becomes
\begin{equation}\label{bevolve:1}
\frac{\partial b}{\partial\lambda}=-2+b''+b'f'-\frac{\alpha}{2}P(r,\lambda)
\end{equation}
where 
\begin{equation}\label{p:1}
P(r,\lambda)=\frac{2}{b}-\frac{b^{'2}}{b^{2}}+\frac{b^{'4}}{4b^{3}}-\frac{b^{'2}b^{''}}{2b^{2}}+\frac{b^{''2}}{2b},
\end{equation}
then, after derivation with respect to radial coordinate, equation (\ref{bevolve:1}) becomes
\begin{equation}\label{minimummod:1}
\frac{\partial b'}{\partial\lambda}=b'''+b''f'+b'f''-\frac{\alpha}{2}\frac{\partial}{\partial r}P(r,\lambda).
\end{equation}

Now, let us assume that $b'>0$ for all $r<r_{a}$, we also assume that at $r=r_{a}$ it appears a minimal surface, therefore $b'(r_{a})=0$. Due to the fact that $b'$ decreased to zero we have $\frac{\partial b'}{\partial\lambda}(r_{a})<0$. On the other side, we have $b''(r_{a})=0$ (since $r_{a}$ is a minimum of $b'$ ), and also $b'''(r_{a})>0$. With all this, and using  (\ref{minimummod:1}) we can see that $\frac{\partial b'}{\partial\lambda}>0$ if $\frac{\partial P}{\partial r}\leq 0$, indeed, from eq.(\ref{p:1}) we can see that $\frac{\partial P}{\partial r}(r_{a})=0$. We have found a contradiction with the original statement of non positiveness of $\frac{\partial b'}{\partial \lambda}$. Trapped regions are bounded by minimal surfaces, then the result follows.
\end{proof}

Once we have shown that no apparent horizon\footnote{In the case of time symmetric data they coincide with minimal surfaces} appears because of the evolution under the flow,  we would like to know how a minimal surface evolves under the RG-2 flow. 
\begin{proposition}
Let us consider  an asymptotically flat  spherically symmetric spacetime written in the b-form. Then the area of an apparent horizon decreases under RG-2 flow .
\end{proposition}
\begin{proof}
Let us assume that during the evolution $r_{a}=r_{a}(\lambda)$. This takes into account the case when the location of the horizon changes. Thus, the area $A(r)=4\pi b$ can be made evolve with a general flow:

\begin{equation}
\frac{dA}{d\lambda}=4\pi\frac{db}{d\lambda}=4\pi\frac{\partial b}{\partial \lambda}|_{r=r_{a}},
\end{equation}
using the RGo-2 flow (without the DeTurck term):
\begin{equation}
\frac{\partial b}{\partial \lambda}=\frac{\partial h_{\theta\theta}}{\partial \lambda}=-2R_{\theta\theta}-\frac{\alpha}{2}Rm_{\theta\theta},
\end{equation}
we found
\begin{equation}
\frac{\partial b}{\partial \lambda}=-2(1-\frac{b^{''}}{2})-\frac{\alpha}{2}(8b^{2}-2b^{''2}b^{2}).
\end{equation}
Using the condition(\ref{apparent}) and since $b^{''}< 1$ we get after multiplying the equation by $4\pi$
\begin{equation}\label{ahorizon:1}
\frac{dA}{d\lambda}<-4\pi(1+\alpha b^{2}).
\end{equation}
Now, since $\alpha>0$ we can conclude that the area of an apparent horizon decreases under the RG-2 flow.
\end{proof}
The rate of decreasing is not constant as in the case of Ricci flow. In the flat metric case ($b=1$) the rate is slightly modified by the parameter $\alpha$. We will see the difference between the rates of both flows when $b$ is big enough, and in any case, when there is a minimal surface present the RG-2 flow make it disappear. We can conclude that if the horizon persist  then the area goes to zero in a finite $\lambda$ but faster than in the case of the Ricci flow.
In the case when $b\rightarrow 0$ we will approach the singularity, and  since the term corresponding to the RG-2 flow goes to zero ($\alpha b^{2}\rightarrow 0$) we recover the Ricci flow, and therefore, an apparent horizon that remains under RG-2 flow results in a singularity in a finite $\lambda$.

\subsection{Area under RG-2 flow}
Let us consider an asymptotically flat manifold $\Sigma$ with one end at infinity. As before, we also take a closed orientable surface $S\subset\Sigma$ with metric $\gamma_{ij}$, which will be the induced metric of $h_{ab}$ on $S$. We call $K_{ij}$ to the extrinsic curvature of $S$ and $K=K^{ij}\gamma_{ij}$ its trace. If we make evolve  the metric $h_{ab}$ (the metric for $\Sigma$) under an intrinsic flow we should expect that the unit normal vector to $S$ will change together with the flow. Thus, we want to know how the area changes with $\lambda$ while the metric $h_{ab}$ evolves under the RG-2 flow eq.(\ref{renequanew:3}). The particular case for the evolution under Ricci flow was considered in \cite{Samuel:2007ak}.

Following the same notation as in \cite{Poisson} we can define our surface $S$ as the level set of a function $\Phi$ on $\Sigma$. This function $\Phi$ is taken as strictly increasing outward from $S$. The unit normal $n_{a}$ is given by $n_{a}=\frac{\Phi_{a}}{\sqrt{(\Phi\Phi)}}$ and it will depend on the metric. The general variation of the area eq.(\ref{area:1}) is given by  \cite{Samuel:2007ak}:
\begin{equation}\label{areavariation:1}
\frac{dA}{d\lambda}=\frac{1}{2}\int_{S}\sqrt{\gamma}d^{2}x(h^{ab}-n^{a}n^{b})\frac{dh_{ab}}{d\lambda}.
\end{equation}
The expression for the evolution of the area given in eq (\ref{areavariation:1}) can be used for any kind of intrinsic flow, for the case at hand we just have to replace $\frac{dh_{ab}}{d\lambda}$ by using eq.(\ref{renequanew:3}). We should get the already known expression for the Ricci flow plus a term of higher order curvature multiplied by $\alpha$, then in the limit when $\alpha$ goes to zero we must recover the Ricci flow case.\\
\begin{theorem}\label{theorem1}
The evolution of the area $A$ of the surface $S$ under the RG-2 flow satisfies:
\begin{equation}\label{ineq:1}
\frac{dA}{d\lambda}\leq -\frac{1}{4}C(S)-\frac{\alpha}{4}C_{R\mathcal{R}}(S),
\end{equation}
where
\begin{equation}
C_{R\mathcal{R}}=\int_{S}\sqrt{\gamma}d^{2}x R(\mathcal{R}+\frac{1}{4}K^{2}),
\end{equation} 
and $C(S)$ given by (\ref{compactness}), with $\mathcal{R}$ being the scalar curvature of the surface $S$, and $R$ the scalar curvature of the ambient manifold.
\end{theorem} 
\begin{proof}
The variation of the area of $S$ under RG-2 flow is given by:
\begin{equation}\label{areavar:1}
\frac{dA}{d\lambda}=\frac{1}{2}\int_{S}\sqrt{\gamma}d^{2}x(-R+n^{a}n^{b}R_{ab}-\frac{\alpha}{2}Rm^{2}+\frac{\alpha}{2}n^{a}n^{b}Rm^{2}_{ab}).
\end{equation}
Here we have to be careful, in the case of the Ricci flow the existence of an ancient solution was established for all $\lambda$,  however for the RG-2 flow there is only a particular zone where the flow is weakly parabolic and we can  assure the existence of the solution only for a finite time.
Using the Gauss-Codazzi equations 
\begin{align}
n^{a}n^{b}R_{ab}&=\frac{1}{2}\left(R-\mathcal{R}-(K_{ij}K^{ij}-K^2)\right),
\end{align}
in (\ref{areavar:1}) and the identity
\begin{equation}
(K_{ij}K^{ij}-K^{2})=(K^{ij}-\frac{1}{2}K\gamma^{ij})(K_{ij}-\frac{1}{2}K\gamma_{ij})+\frac{1}{4}K^{2},
\end{equation}
we arrive to
\begin{equation}\label{areavariation:2}
 \frac{dA}{d\lambda}=-\frac{1}{4}\int_{S}\sqrt{\gamma}d^{2}x(2\mathcal{R}-K^{2})+\\
\int_{S}\sqrt{\gamma}d^{2}x(R+K_{ij}K^{ij}-\alpha Rm^{2}+\alpha n^{a}n^{b}Rm^{2}_{ab}).
\end{equation}
It would be helpful to use some type of Gauss-Codazzi equations for the higher order terms, however there is a way of simplifying the expression term by term. The Kretschmann scalar $Rm^{2}$ in $d-$dimensions can be written as:
\begin{equation}
R_{abcd}R^{abcd}=C_{abcd}C^{abcd}+\frac{4}{d-2}R_{ab}R^{ab}-\frac{2}{(d-1)(d-2)}R^{2}
\end{equation}
where $C_{abcd}$ is the Weyl tensor (it vanishes in three dimensions). Moreover, in three dimensions, the term $Rm^{2}_{ab}$  can be written as a function of the Ricci tensor and the scalar curvature, see eq.(\ref{fixedpointdec}). After some massaging we arrive to:

\begin{eqnarray}
\frac{dA}{d\lambda}&=&-\frac{1}{4}C(S)
+\frac{\alpha}{4}\int_{S}\sqrt{\gamma}d^{2}x(-2\eta^{a}\eta^{b}R_{a}^{\,\,\sigma}R_{\sigma b}\nonumber\\
&&
-R(\mathcal{R}+(K^{ij}-\frac{1}{2}K\gamma^{ij})(K_{ij}-\frac{1}{2}K\gamma_{ij})+\frac{1}{4}K^{2})\nonumber\\
&&
-2|Rc|^{2}-2R^{2}.
\end{eqnarray}
Rearranging some terms and using the identity
\begin{equation}
\eta^{b}R_{ab}=\nabla K_{a}-\nabla^{c}K_{ac},
\end{equation}
we can write
\begin{eqnarray}\label{areavariation:3}
\frac{dA}{d\lambda}&=&-\frac{1}{4}C(S)
+\frac{\alpha}{4}\int_{S}\sqrt{\gamma}d^{2}x\{-2|Rc|^{2}-2R^{2}\nonumber\\
&&
-2(\nabla K_{a}-\nabla^{c}K_{ac})(\nabla K^{a}-\nabla^{c}K_{c}^{a})\nonumber\\
&&
+R(K^{ij}-\frac{1}{2}K\gamma^{ij})(K_{ij}-\frac{1}{2}K\gamma_{ij})-R(\mathcal{R}+\frac{1}{4}K^{2})\}.
\end{eqnarray}

All terms are of definite sign then the result follows. 
\end{proof}

Theorem \ref{theorem1} is one of our main results. Inequality (\ref{ineq:1}) is saturated by the Schwarzschild solution, and therefore it refines the result already obtained in \cite{Samuel:2007ak}. Thus, we have the following corollary:
\begin{corollary}
The evolution of the area $A$ of the surface $S$ under the RG-2 flow satisfies:
\begin{equation}\label{ineq:2}
\frac{dA}{d\lambda}\leq -\frac{1}{4}C(S).
\end{equation}
\end{corollary}
\begin{proof}
The quantity $C_{R\mathcal{R}}$ in (\ref{ineq:1}) is positive, then the result goes directly  .
\end{proof}
The corollary shows that the rate of change of the area under the RG-2 flow is bigger than the rate under the Ricci flow, and therefore, the area of a compact surface shrinks to a point faster than in the case of Ricci flow. As we consider more loops the rate of change will increase, therefore we conjecture that the term $\frac{1}{4}C(S)$  represents an upper bound for all loops. 
\begin{corollary}
The evolution of the area $A$ of the surface $S$ under the RG-2 flow satisfies:
\begin{equation}\label{ineq:3}
\frac{dA}{d\lambda}\leq -\frac{\alpha}{4}C_{R\mathcal{R}}(S)
\end{equation}
\end{corollary}
\begin{proof}
The quantity $C_{S}$ in (\ref{ineq:1}) is positive, then the result goes directly  .
\end{proof}
The inequality (\ref{ineq:3}) is not easy to interpret. It can also be proved considering the evolution of $A$ under the RGo-2 flow. The quantity $C_{R\mathcal{R}}$ is some kind of ``compactness", however it includes the curvature  $R$ of the ambient manifold. It cannot be saturated by the Schwarzschild metric, and as we increase the number of loops more terms will appear, and therefore more bounds. However, none of those bounds  will be saturated by the Schwarzschild solution. In order to get an insight about this higher order term let us see what happens when our surface $S$ is a sphere. In this case we found that the variation for the area eq (\ref{area:1}) under RG-2 flow is:
\begin{equation}
\frac{dA}{d\lambda}=-2\pi\int_{S}\sqrt{\gamma}d\theta d\phi R-\frac{1}{4}C-\frac{\alpha}{2}\left(\frac{b^{'2}}{2b}-b^{''}\right)^{2},
\end{equation}
where $C$ is the compactness defined in eq.(\ref{compactness}).
Since $R>0$ we arrive to 
\begin{equation}\label{areavariation:5}
\frac{dA}{d\lambda}\leqslant-\frac{1}{4}C-\frac{\alpha}{2}\left(\frac{b^{'2}}{2b}-b^{''}\right)^{2}.
\end{equation}
The inequality (\ref{areavariation:5}) can be written as a function of the Hawking mass 
\begin{equation}
\frac{dA}{d\lambda}\leqslant-16\pi^{3/2}\frac{M_{H}}{\sqrt{A}}-\frac{\alpha}{2}\left(\frac{b^{'2}}{2b}-b^{''}\right)^{2}.
\end{equation}

We can see that the presence of the term with $\alpha$ pushes down the bound for the area rate change under the flow. As we expected the convergence to a singularity will be faster than in the case with the Ricci flow. Since all terms have a definite sign we also have the bound
\begin{equation}\label{areavariations:1}
\frac{dA}{d\lambda}\leqslant-\frac{\alpha}{2}\left(\frac{b^{'2}}{2b}-b^{''}\right)^{2}.
\end{equation}

Due to the fact that when $b\rightarrow 0$ we approach the singularity, we can see from (\ref{areavariations:1}) that the part multiplying $\alpha$ goes to infinity, making the surface shrink to a point very fast.
The new quantity $C_{R\mathcal{R}}$ involves not only the geometric information concerning the surface $S$ but also the information about curvature of the ambient manifold, but as we see in the spherical case it can be recast as a positive quantity multiplied by $\alpha$.
As before, we conjecture that after the second loop contribution every quantity is of definite sign, therefore the inequality
\begin{equation}
\frac{dA}{d\lambda}\leqslant-16\pi^{3/2}\frac{M_{H}}{\sqrt{A}},
\end{equation}
will be satisfied for all loops. 

\subsection{Hawking mass under RG-2 flow}
Here we develop a formula for the evolution of the Hawking mass given in (\ref{Hawking:1}) under the RG-2 flow. It is clear that the evolution of Hawking mass can be computed using the variation of the area eq. (\ref{area:1}) and the variation of the compactness eq.(\ref{compactness}). 
Since the  term $\int d^{2}x\sqrt{\gamma}\mathcal{R}$ is a topological invariant it vanishes after differentiation, then the general evolution of compactness under RG-2 flow is given by \cite{Samuel:2007ak}:

\begin{eqnarray}\label{cvariation}
\frac{dC}{d\lambda}
& = & 
-{\displaystyle\int}
dA K\left\{h^{ab}n^{c}\nabla_{c}\left(\frac{dh_{ab}}{d\lambda}\right)
-2\nabla_{a}\left(\frac{dh^{ab}}{d\lambda}n_{b}\right)
+n^{a}\nabla_{a}\left(\frac{dh^{cd}}{d\lambda}n_{c}n_{d}\right)\right\}
\nonumber\\
& &
-{\displaystyle\int}
dA\frac{K^{2}}{2}\left(\frac{dh^{ab}}{d\lambda}n_{c}n_{d}
+h^{ab}\frac{dh^{ab}}{d\lambda}
\right).
\end{eqnarray}

The equation (\ref{cvariation}) works for any intrinsic flow, here we are interested in the RG-2 flow, and as usual we should recover the Ricci flow case when $\alpha\rightarrow 0$.
After substitution of (\ref{renequanew:3}) into (\ref{cvariation}) we found:

\begin{eqnarray}\label{eq:dclambda2}
\frac{dC}{d\lambda}
& = & 
{\displaystyle\int}
dA K\left\{
h^{ab}n^{c}\nabla_{c}\left(2R_{ab}
+\frac{\alpha}{2}Rm^{2}_{ab}\right)
+2\nabla_{a}\left(2R^{ab}n_{b}
+\frac{\alpha}{2}Rm^{2\,\,ab}n_{b}\right)
\right.
\nonumber\\
& & 
\left.
-n^{a}\nabla_{a}\left(2R^{cd}n_{c}n_{d}
-\frac{\alpha}{2}Rm^{2\,\,cd}n_{c}n_{d}\right)\right\}
\nonumber\\
& &  
+{\displaystyle\int}
dA K^{2}\left\{R
+n^{a}n^{b}R_{ab}
+\frac{\alpha}{4}\left(R_{m}^{2\,\,\,cd}n_{c}n_{d}
+R_{m}^{2}\right)
\right\}.
\end{eqnarray}

After some simplifications we obtain
\begin{eqnarray}
\frac{dC}{d\lambda}
& = & 
{\displaystyle-\int}
2dA K\left\{2R^{ab}\nabla_{a}n_{b}-n^{a}\nabla_{a}(R^{ab}n_{a}n_{b})\right.
\nonumber\\
&&
\left.
-\frac{\alpha}{2}\left(\frac{1}{2}n^{c}\nabla_{c}Rm^{2}-\nabla_{a}(Rm^{2ab}n_{b})+\frac{1}{2}n^{a}\nabla_{a}Rm^{2ab}n_{a}n_{b}\right)\right\}
\nonumber\\
&&
+{\displaystyle\int}
dA K^{2}\left\{R
+n^{a}n^{b}R_{ab}
+\frac{\alpha}{4}\left(R_{m}^{2\,\,\,cd}n_{c}n_{d}
+R_{m}^{2}\right)
\right\}.
\end{eqnarray}
Once we have calculated the variation of the area and the variation of the compactness we can calculate the variation of Hawking mass as follows:
\begin{equation}
\frac{dM_{H}}{d\lambda}=\frac{1}{64\sqrt{A\pi^{3}}}\left(\frac{1}{2}\frac{dA}{d\lambda}C+A\frac{dC}{d\lambda} \right).
\end{equation}
We know that compactness is not monotonic under the Ricci flow, adding a term that can be very big near a singularity could change this fact, however we saw that the extra terms that appear because of the new flow are of definite sign,  therefore, they cannot make compactness monotonic under RG-2 flow.

\section{Conclusions and final comments}

We have studied the behaviour of the two loop renormalization group flow in the context of general relativity. Focusing in asymptotically flat spaces we have generalized the results obtained for the Ricci flow in \cite{Samuel:2007ak}, being the inequality (\ref{ineq:1}) one of the main results. The inequality (\ref{ineq:2})  was known to be satisfied when we make evolve a closed surface under the Ricci flow, but now we know that it is also satisfied when the evolution is under RG-2 flow, and there are strong reasons to believe that the inequality is going to be satisfied for all loops. Some results for the Ricci flow are also extended to the RG-2 flow, one of them is about the type of surface that we make evolve. If the surface $S$ is an outermost horizon we can identify it with the area of a black hole horizon, then, our result shows that an outermost horizon always shrink to a point when evolved by the RG-2 flow, but only in the zones where sectional curvatures satisfy $1+\alpha K_{ab}>0$. It implies that the black hole entropy-as it is related with the area- is monotonously decreasing under RG-2 flow. Although, we would expect that the entropy increases along the flow it is not clear how we can relate the irreversibility and the entropy in the case of the RG-2 flow.\\
We have made evolve a closed surface under RG-2 flow, and since the entanglement entropy for the two sub-systems separated by the surface is proportional to the area of the surface -it also depends on the UV cutoff- our results can be used for studying the evolution of entanglement entropy in a similar way that has been done with the Ricci flow in \cite{Solodukhin:2006ic}. We expect similar results although there is not a clear way in how to approach the study with the new flow.
There is a connection  between entanglement entropy and thermal entropy, and in some way, the evolution results obtained for entanglement entropy will be related to those for the thermal one. The analogy is not perfect, in the UV limit both entropies are not related at all, however in the IR limit the entanglement entropy is made of two parts: the thermal entropy corresponding to the Bekenstein-Hawking entropy of the dual black hole geometry  and the remaining quantum entanglement near the entangling surface, which is small compared with the thermal entropy in the IR region\cite{Kim:2016jwu}. 
Extending the calculations from two loops to higher loops is straightforward, although it is going to be cumbersome. \\ We found another inequality which is not saturated by the Schwarzschild solution but it could give important information about the evolution relation between the ambient manifold and a surface defined on it. It will be interesting to see how this quantity evolve under RG-2 flow. Studying the behaviour of the metric under the RG-2 flow  near the points with higher curvatures could give a clue about  the singularities developed by the flow and how to treat them.\\ Global quantities such as the ADM mass have been studied in the context of the Ricci flow \cite{Dai:2005}\cite{Oliynyk:2006nr}, and it would be interesting to know how the effects of higher curvature terms change the results \cite{Lasso-Andino:1}.
We also have discussed about the gradient formulation of the RG-2 flow, this formulation is an open mathematical problem and we think that higher order gravities will give a clue in finding the formulation, a good start is the Einstein Gauss-Bonnet  gravity coupled to a scalar field, clearly it will involve a non linear heat equation. 
\newpage

\section{Appendix}
\appendix
\section{W-entropy and the RG-2 flow}

The intrinsic flows such as Ricci flow and RG-2 flow provide an analytic way of deforming a metric by smoothing the zones with higher curvatures. In a similar way that the heat equation tends to smooth (homogenize) the temperature of a surface, geometric flows homogenize the curvature. This analogy goes far beyond that. In Perelman's work, the introduction of a $\mathcal{W}$-entropy was  pivotal, and hand by hand with the gradient formulation of the flow it became an important tool for the proof of Thurston`s geometrization conjecture. Recently, it has been shown that the $W-$entropy for the Ricci flow  has a clear interpretation as the derivative of the difference  between the Boltzman-Shannon entropy of the heat kernel measure and the Boltzman-Shannon entropy of the gaussian   measure\cite{Li1}\cite{Li2}\cite{Li3}. We would like to define a  $\mathcal{W}$-entropy  and study its behaviour under RG-2 flow. The gradient formulation for the RG-2 flow is not known, although a recent work has some results in this direction \cite{Carfora:2018}. Here we will argue in favour of its existence and give some clues about the construction. A good start is the gradient formulation of the Ricci flow. We will now focus on compact three dimensional manifolds and give a  brief summary here.\\
Let $M$ be a 3-dimensional compact Riemannian manifold, let us define the following functional
\begin{equation}
\mathcal{F}(h_{ab},f)=\int_{M}\left(R+|\nabla f|^{2}\right)e^{-f}dv.
\end{equation}
Here $h_{ab}$ denotes the Riemannian metric defined on $M$, $f\in C^{\infty}(M)$ and $R$ is the scalar curvature of $h_{ab}$. The integration measure is $dv$ and it is constrained such that $d\mu=e^{-f}dv$ remains fixed along the flow.
In \cite{Perelman:2006un} Perelman showed that the gradient flow with respect to the metric in $M\times C^{\infty}(M)$ is given by the following system of partial differential equations:
\begin{eqnarray}
\frac{\partial h_{ab}}{\partial\lambda}&=&-2(R_{ab}+\nabla_{a}\nabla_{b}f)\\
\frac{\partial f}{\partial\lambda}&=&-\Delta f-R
\end{eqnarray}
Together with the functional $\mathcal{F}$ Perelman defined a functional $W$ called entropy:
\begin{equation}
W(h_{ab},f,\lambda)=\int_{M}\left(\tau(R+|\nabla f|^{2})+f-n\right)\frac{e^{-f}}{(4\pi \tau)^\frac{n}{2}}dv
\end{equation}
where $\tau>0$, and as before it is imposed the constraint
\begin{equation}
\int_{M}(4\pi\tau)^{-\frac{n}{2}}e^{-f}dv=1.
\end{equation}
He gave a remarkable proof. Assuming that $(h_{ab},f,\tau)$ satisfy
\begin{eqnarray}
\frac{\partial h_{ab}}{\partial\lambda}&=&-2 R_{ab},\label{Ricci:1}\\
\frac{\partial f}{\partial\lambda}&=&-\Delta f+|\nabla f|^{2}-R+\frac{n}{2\tau},\label{Ricci:2}\\
\frac{\partial\tau}{\partial\lambda}&=&-1\label{Ricci:3},
\end{eqnarray}
then the variation of the $\mathcal{W}$-entropy under Ricci flow is given by
\begin{equation}\label{entropy:var}
\frac{d}{d\lambda}\mathcal{W}(h_{ab},f,\tau)=2\int_{M}\tau\left|R_{ab}+\nabla_{a}\nabla_{b}f-\frac{h_{ab}}{2\tau}\right|^{2}\frac{e^{-f}}{(4\pi\tau)^{\frac{n}{2}}}dv.
\end{equation}
The $\mathcal{W}$-entropy is monotonic increasing in $\lambda$, except for the shrinking solitons $(M,h_{ab},f)$ that by definition satisfy
\begin{equation}
R_{ab}+\nabla_{a}\nabla_{b}f=\frac{h_{ab}}{2\tau}.
\end{equation}
According to the Perelman's probability interpretation, if there is a partition function $Z_{\beta}$  in a canonical ensemble where  the density of states measure is given by $g(E)dE$ then the partition function $Z=\int_{\mathbb{R}^{+}}e^{-\frac{1}{\tau}E}g(E)dE$ is defined by
\begin{equation}\label{partitionf}
\log Z_{\beta}=\int_{M}\left(-f+\frac{n}{2}\right)d\mu,
\end{equation}
where 
\begin{equation}
d\mu=(4\pi\tau)^{-\frac{n}{2}}e^{-f}dv,
\end{equation}
then we can calculate 
\begin{equation}
S=\frac{\partial}{\partial \tau}(\tau \log Z_{\beta}),
\end{equation}
where we have set $\tau=\beta^{-1}$. Morever, 
\begin{equation}
\langle E\rangle=-\frac{\partial}{\partial\beta}\log Z_{\beta}=-\tau^{2}\int_{M}\left(R+|\nabla f|^{2}-\frac{n}{2\tau}\right)\frac{e^{-f}}{(4\pi\tau)^{\frac{n}{2}}}dv,
\end{equation}
\begin{equation}
\langle(E-\langle E\rangle)^{2}\rangle=\frac{\partial^{2}}{\partial \beta^{2}}\log Z_{\beta}=2\tau^{4}\int_{M}\left|R_{ab}+\nabla_{a}\nabla_{b}f-\frac{h_{ab}}{2\tau}\right|^{2}\frac{e^{-f}}{(4\pi\tau)^{\frac{n}{2}}}dv,
\end{equation}
\begin{equation}
S=-\int_{M}\tau \left(R+|\nabla f|^{2}-f-n\right)\frac{e^{-f}}{(4\pi\tau)^{\frac{n}{2}}}dv.
\end{equation}
Then
\begin{equation}
W=-S.
\end{equation}

Since the apparition of Perelman works there has been a lot of  efforts for understanding the entropy in different contexts. In \cite{Li2}, starting with the heuristic interpretation of the $\mathcal{W}$-entropy given by Perelman in the context of statistical mechanics, Li gave a probabilistic interpretation of this entropy for the heat equation $\frac{\partial u}{\partial\lambda}=Lu$ on complete Riemannian manifolds with the assumption of a weighted volume measure.
Assuming that $(h_{ab},f)$ is the solution of the system of equations given by (\ref{Ricci:1}), (\ref{Ricci:2}) and (\ref{Ricci:3}) then:
\begin{eqnarray}\label{partition}
\log Z_{\beta}&=\int_{M}\left(\frac{n}{2}-f\right)d\mu
&= \frac{n}{2}\left(1+\log (4\pi\tau)\right)-H,
\end{eqnarray}
where
\begin{equation}
H=\int_{M}u \log u dv=\int_{M}\left(f+\frac{n}{2}\log(4\pi\tau)\right)\frac{e^{-f}}{(4\pi \tau)^{n/2}}.
 \end{equation}
This $H$ is called the Boltzman-Shannon entropy of the measure $udv$ with
\begin{equation}
u=\frac{e^{-f}}{(4\pi\lambda)^{n/2}}.
\end{equation}
Moreover, it is known that the Boltzman-Nash-Shanon entropy of the Gaussian heat kernel measure on $\mathbb{R}^{n}$ is given by 
\begin{equation}
H(\gamma^{n}_{\tau})=-\frac{n}{2}\left(1+\log(4\pi\tau)\right).
\end{equation}
We can write the $\mathcal{W}$-entropy as
\begin{equation}
W=\frac{d}{d\tau}(\tau\left(H(\gamma_{\tau}^{n})-H\right))
\end{equation}
Defining $F=-\tau \log Z$ the entropy becomes $\mathcal{W}=-\frac{\partial F}{\partial \tau}$, $F$ is called the Helmholtz free energy function. 

As the reader should have noticed, the interpretation in the context of statistical mechanics is completely based in the definition of the partition function $Z$, once we have defined it all "macroscopic" quantities can be calculated. There is no physical interpretation for the partition function defined in (\ref{partitionf}).\\
In section \ref{solutions} we presented the definition of steady gradient solitons  where the $\alpha$ term is the extra contribution because of the higher curvature terms, therefore we expect that the evolution of a $W$-2-entropy under the RG-2 flow should be of the form
\begin{equation}
\frac{d}{d\lambda}\mathcal{W}_{2}\approx\int_{M}\left|R_{ab}+\frac{\alpha}{2}Rm_{ab}^{2}+\nabla_{a}\nabla_{b}f-\frac{h_{ab}}{2\tau}\right|^{2}dm,
\end{equation}  
where  $dm$ is a measure satisfying $\int_{M}dm=1$. In this case the $W_{2}$ would be monotonous under the RG-2 flow.  Here we need to specify the weighted measure $dm$ which for the case of the Ricci flow was $udv$ with $u$ given by the heat kernel function (a solution of the conjugated linear heat equation).
Similarly, the functional $\mathcal{F}$ should include scalar functions built by higher order curvature terms, namely  $R^{2}$,$R_{ab}R^{ab}$,$R_{abcd}R^{abcd}$.\\
Nobody knew that the Ricci flow was a gradient flow until Perelman found the formulation, but in the context of physics, and since the Ricci flow is the first order approximation of the RG flow for the $\sigma$-model it was suspected that it was the case. Similarly, we expect that  the RG-2 flow has also  a gradient formulation.\cite{Lasso-Andino:1} \\

Finally, there are some ways to explore the relationship between the $W$ entropy and the entropy of a black hole. In \cite{Samuel:2007zz} the authors applied some of the Perelman ideas to the General relativity scenario, looking for a relationship between the entropy of a black hole with the $\mathcal{W}$-entropy. They studied the fixed points of the Ricci flow and showed that Schwarzschild can not be a fixed point of the flow, and therefore the $\mathcal{W}$-entropy was not related to the Beckestein-Hawking entropy. They proposed a modified flow together with an evolution equation for $f$
\begin{eqnarray}\label{riccisol}
\frac{\partial h_{ab}}{\partial \lambda}&=&-2f(R_{ab}-g h_{ab})+2D_{a}D_{b}f,\\
\frac{\partial f}{\partial\lambda}&=&D^{2}f+g f.\label{heat:1}
\end{eqnarray}
The Schwarzschild-AdS is a Ricci soliton for (\ref{riccisol}) when $f$ is given by
\begin{equation}\label{soliton}
f=\left(1-\frac{2M}{r}-\frac{g}{3r^{2}}\right)^{\frac{1}{2}}.
\end{equation}
The Schwarzschild solution is not a fixed point of the Ricci flow neither of the RG-2 flow. We could follow a similar procedure for the Rg-2 flow, and propose a variation of the RG-2 flow
\begin{equation}\label{newflow}
\frac{\partial h_{ab}}{\partial \lambda}=-2f(R_{ab}-g h_{ab})+\frac{\alpha}{2}f Rm_{ab}^{2}+2D_{a}D_{b}f
\end{equation}

The equation (\ref{soliton:1}) has to be satisfied for the solitons at the beginning of the flow, and it will be satisfied for any finite time $T$. However, we know that coupling the eq. (\ref{heat:1}) to \ref{newflow} does not work. We need to provide a nonlinear "heat" equation in order to make (\ref{soliton}) a soliton \cite{Lasso-Andino:1}.

\section*{Acknowledgments}
This work has been partially supported by Spanish  Research  Agency  (Agencia  Estatal  de  Investigaci\'{o}n) through the grant IFT Centro de Excelencia  Severo  Ochoa  SEV-2016-0597. The work of OLA was also supported by a scholarship of the Ecuadorian Secretary of Science, Technology and Innovation.

\appendix



\end{document}